# DesignCon 2008

# Challenges in implementing DDR3 memory interface on PCB systems: a methodology for interfacing DDR3 SDRAM DIMM to an FPGA

Phil Murray, Altera Corporation

Feras Al-Hawari, Cadence Design Systems, Inc.

## Abstract
Undoubtedly faster, larger and lower power per bit, but just how do you go about interfacing a DDR3 SDRAM DIMM to an FPGA? The DDR3 standard addresses the faster, more bandwidth and lower power per bit need, but it introduces new design challenges in addition to challenges introduced by DDR2 ODT, slew rate derating, etc. The DDR3 fly-by topology requirement means customers designing DDR3 memories must now account for write leveling and read de-skew on the PCB. This paper will cover modeling, simulation, and physical layout approaches required to meet JEDEC-defined termination and tight timing requirements for designing DDR3 memory interfaces on PCB systems.

## Author Biographies
### Phil Murray, Altera Corporation
Phil Murray joined Altera in December 2000 as a high-speed design engineer working in the component applications' high-speed board development group. Mr. Murray's primary responsibilities include designing checkout boards, demonstration boards, and development kits for Altera for customer purchases. Mr. Murray has over 15 years of industry experience, including seven years in communications design. Mr. Murray graduated with his BSEE from San Diego State University.

### Feras Al-Hawari, PhD, Cadence Design Systems, Inc.
Dr. Al-Hawari is a senior member of consulting staff in the Allegro PCB SI high-speed R&D group. Dr. Al-Hawari joined Cadence Design Systems in 1995. Recently, Dr. Al-Hawari architected and automated the post-layout analysis flow for source synchronous interfaces, and implemented methods for S-parameter as well as multi-conductor transmission line simulation. Dr. Al-Hawari's research interests include signal integrity, PCB design and analysis, circuit simulation, and modeling of I/O devices and interconnects, in addition to parallel and network computing. Dr. Al-Hawari has a PhD in electrical engineering from Northeastern University, MS in computer engineering from Florida Institute of Technology and a BS in electrical and computer engineering from Jordan University of Science and Technology.

# Introduction

Some of the challenges designers face when implementing a DDR2 DIMM interface are alleviated by the DDR3 DIMM architecture, but there are still some challenges for designers to overcome when implementing DDR3 DIMM interfaces. DDR3 is faster than DDR2 and uses a lower power rail. However, the combination of a lower power rail and faster speed introduces a need for tighter noise margins and less SSN. The fly-by termination architecture of the DDR3 DIMM reduces the number of simultaneous switching signals but causes flight-time skew, which can be up to two cycles across the DIMM. Therefore, a write-and-read-"leveling" feature was defined in DDR3 memories to enable controllers to compensate for this skew by adjusting the timing per byte lane.

In addition to understanding the new DDR3 features, the designers still need to tackle the various challenges that are common in both DDR2 as well as DDR3 memory interfaces. Thus, an effective design methodology geared towards verifying complex DDR3 designs should also address issues such as analyzing the interface for all possible DRAM and controller on-die termination (ODT) configurations, validating the design for all I/O buffers in fast/typical/slow modes (i.e., process, voltage, and temperature (PVT) variations), simulating all memory/controller read/write combinations, and adjusting the setup and hold times based on the slew rates of the data and strobe signals as well as user-defined derating tables.

In this paper, we discuss the migration from DDR2 to DDR3. We explain the new read-and-write-leveling feature that is introduced in the JEDEC DDR3 specification. Then we present a methodology to aid in the design and verification of DDR3 interfaces, as well as show how to set up the interface and define signal associations to automate the simulation of various read-write cycle combinations with different ODT configurations. A pre-layout solution space exploration phase is proposed to develop constraints that drive the system layout process. We explain why slew rate derating is required to meet the desired timing requirements, and finally, discuss how to ensure that the timing and noise margins are not affected by reflection and crosstalk noise in the byte lanes.

# Migration to DDR3

DDR3 is the latest generation of the DDR SDRAM technology and has some distinct advantages over its predecessor DDR2. Some of the challenges designers faced with implementing a DDR2 DIMM interface are alleviated by the DDR3 DIMM architecture. The DDR2 data rate range of 400-800 Mbps doubled with DDR3 to 800-1600 Mbps. DDR3 uses less power because its voltage lowered from 1.8V to 1.5V. The DDR3 DIMM has less loading and less SSN than the DDR2 DIMM due to its fly-by termination architecture. In addition, the DDR3 DIMM has termination for its command/address/control bus located on the DIMM, which increases board real estate by removing the need for external termination. Like DDR2, DDR3 has ODT programmability for its DQS byte lanes. In other words, it has the ability to dynamically turn off its termination during reads and dynamically turn on its termination during writes. For this feature to have its full effect, it is desirable for the FPGA to have this same feature.

These new breakthroughs, however, create new challenges. Increasing the performance and decreasing the voltage causes the need for tighter noise margins, which affect the amount of allowable SSN and produce a smaller data capture window. Proper simulation tools are essential in setting up a DDR3 system to have minimal SSN. A FPGA with adjustable drive strengths can also help to minimize SSN. The DDR3 DIMM's fly-by termination architecture causes flight-time skew between the clock and data strobes at every DRAM as the clock and command/address/control signals traverse the DIMM. These flight-time skews can be accounted for by the DDR3 DIMM's write-and-read-leveling feature.

## DDR3 Write and Read Leveling

DDR3 write and read leveling effectively adds delays to each DQS group so they arrive at each device at the same time regardless of the flight-time skew across the DIMM. The DDR3 command/address/control and clocks arrive at the DIMM at each SDRAM device sequentially. This eliminates trace stubs, but causes flight-time delay between each DRAM clock and strobe. For DDR3 write leveling, the controller needs to launch the DQS groups at separate times to coincide with the memory clock arriving at each device on the DIMM. Before the data is presented to the DDR3 DIMM, the controller needs to setup the appropriate delays (see Figure 1) so that the data will arrive at the correct time to each DDR3 DRAM device on the DIMM. The command/address/control bus and clock from the DDR3 controller to the DDR3 DIMM enters the DIMM through the middle, and then connects sequentially to each DDR3 DRAM device in a daisy-chain pattern.

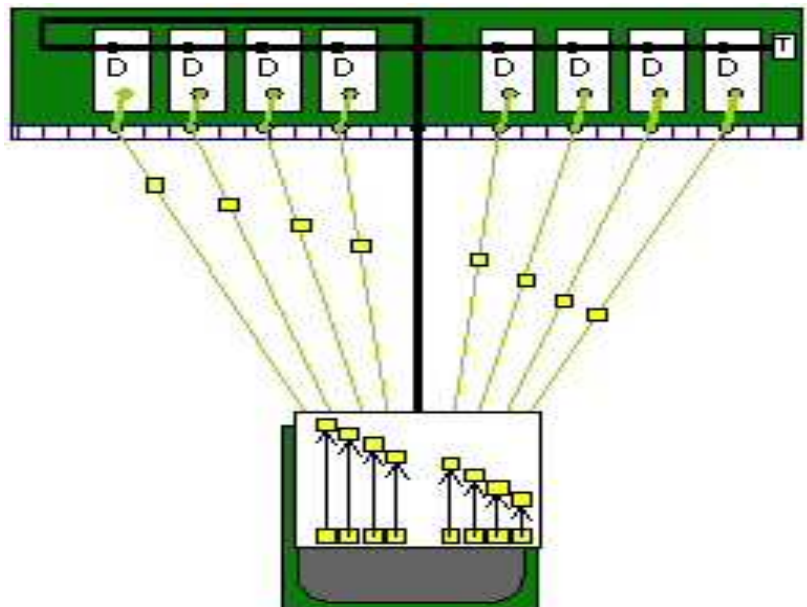

Figure 1: DDR3 DIMM controller using write leveling

The delay from the DDR3 DRAM device on the far left to the device on the far right can be as much as 1.6 ns. The controller's write leveling needs to compensate for this flight-time across the module. The tDQSS (DQS, DQS# rising edge to CK, CK# rising edge) at the DDR3 DRAM device needs to be kept to within 0.25 times the clock period. The controller needs to adjust its DQS and CK signals through feedback from each of the DDR3 DRAM devices while in write-leveling mode.

Like DDR3 write leveling, DDR3 read leveling needs the controller to delay each DQS group coming from the DIMM at individual intervals so that all the DQS groups arrive at the controller at the same time. Read data arrival time at the controller could spread up to two CAS latencies. The first device the address sees on the DIMM is the DRAM device that is located furthest from the termination resistor (see Figure 2). The controller must add the most delay to this device's DQS group to match the DQS group flight-time from the DRAM device that is closest to the termination resistor on the DIMM.

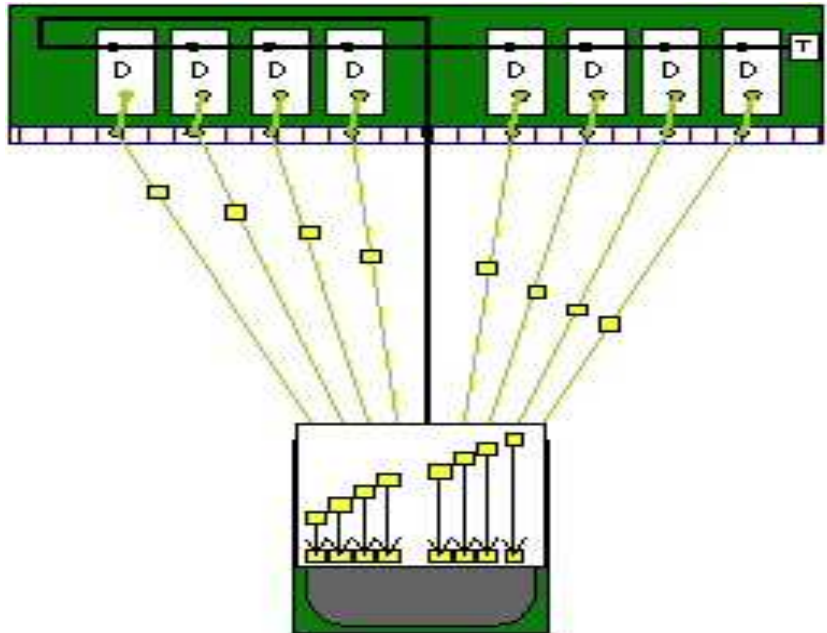
Figure 2: DDR3 DIMM controller using read leveling

# A Methodology for Setup and Analysis of DDR3 Interfaces

In this section, we propose a methodology (i.e., a sequence of steps, shown in Figure 3) to tackle the various DDR3 design challenges. These steps allow the designer to set up the interface, and to define the signal relationships and ODT programmability in order to enable an EDA tool to simulate all memory/controller read/write combinations for all ODT and I/O PVT variations. In addition, we show how read/write leveling can be modeled and simulated. Moreover, we discuss how a tool can adjust the measured setup and hold times based on the slew rates of the data and strobe signals as well as user-defined derating tables.

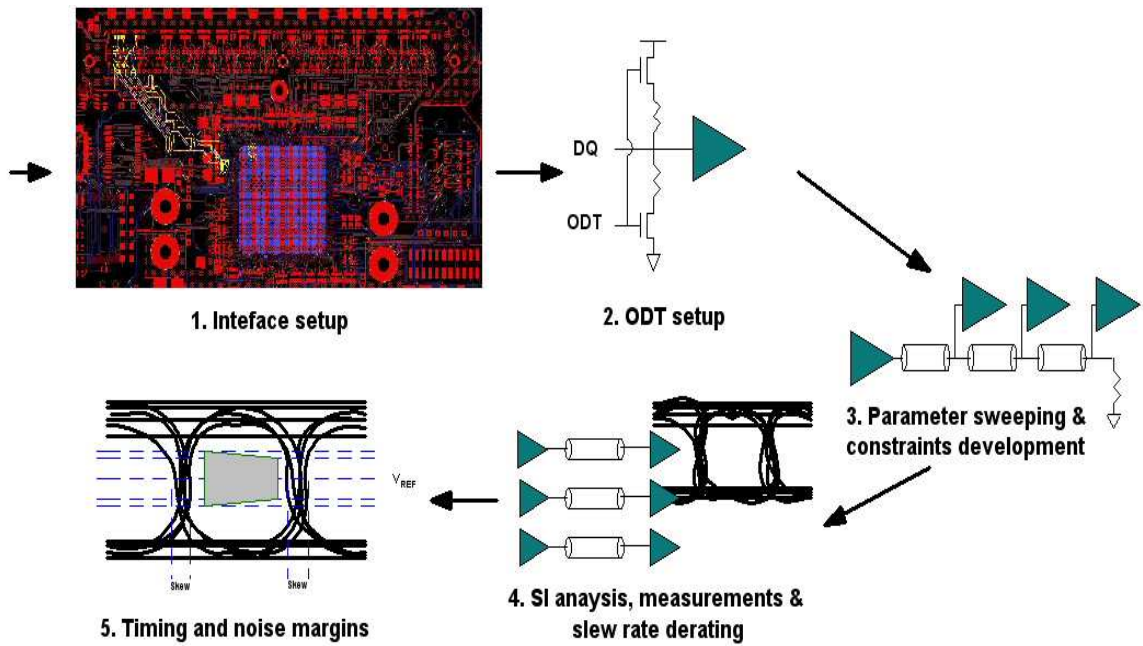

Figure 3: A sequence of steps to design and verify a DDR3 interface

## Interface Setup and Signal Associations

A typical DDR3 interface contains one controller as well as multiple DRAMs distributed on one or more DIMM(s) (see Figure 4(a)). In addition, there are four types of signal groups (data, address/command, control, and clock) interconnecting the controller with the various DRAMs. Moreover, the direction of a signal depends on its group type. For example, an address signal always flows from the controller towards the DRAMs and not vise versa. Thus, defining the buses (i.e., signal groups) in the interface as well as tagging the controller component is required to identify all the permissible driver/receiver combinations to be analyzed.

Based on that, an EDA tool must enable the designer to setup the various DDR3 buses. A bus in this context is an object that contains a group of nets. There are two types of buses in a DDR3 interface: unidirectional (e.g., address bus) and bidirectional (e.g., data bus). On a unidirectional bus, the address/command/control signals only propagate from the controller towards the memory modules. However, on a bidirectional bus the data is read from, or written to, the various DRAMs. Furthermore, the data is latched on the rising or falling edge of a clock or both edges of a strobe when the bus is unidirectional or bidirectional, respectively.

In addition, in a source synchronous interface, a data/address signal is always transmitted along with its corresponding strobe/clock signal to minimize skew. Hence, these signal associations should be captured in order to enable a tool to automatically simulate the required nets and then measure setup and hold times accordingly. To support the various possible signal relationships in a DDR3 interface (see Figure 4(a)), the tool must enable the association of a whole bus to a corresponding clock/strobe signal. In addition, it should allow the association of different lanes or bits of a bus to different clock/strobe signals.

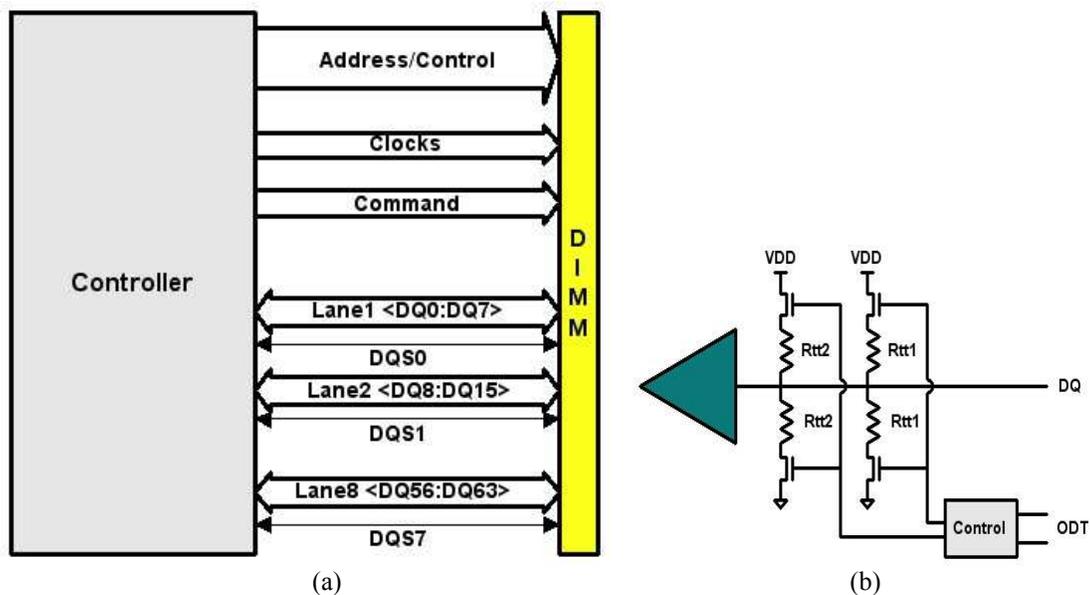

(a) (b)
Figure 4: (a) A 1-DIMM DDR3 interface, and (b) ODT circuitry and control

## On-Die Termination Programmability

DDR3, similarly to DDR2, supports programmable ODT for the data signals in order to suppress reflections and reduce the number of components on the motherboard. The DRAM ODT pin(s) can be utilized to turn the terminations on/off or to select different termination values as needed (see Figure 4(b)). This capability enables the designer to vary the ODT setup of the active and stand-by receivers in such a way that improves signal integrity (SI) in the system. For example, a 2-DIMM system is usually configured during memory reads and writes, as shown in Figures 5(a) and 5(b), respectively. In both configurations, the ODT of the DRAM2 stand-by receiver is always turned on, while the ODT of the controller I/O is only turned on when it is receiving (i.e., during the read cycle). Note that, the best ODT configuration can be determined based on the results of a pre- or post-layout solution space exploration.

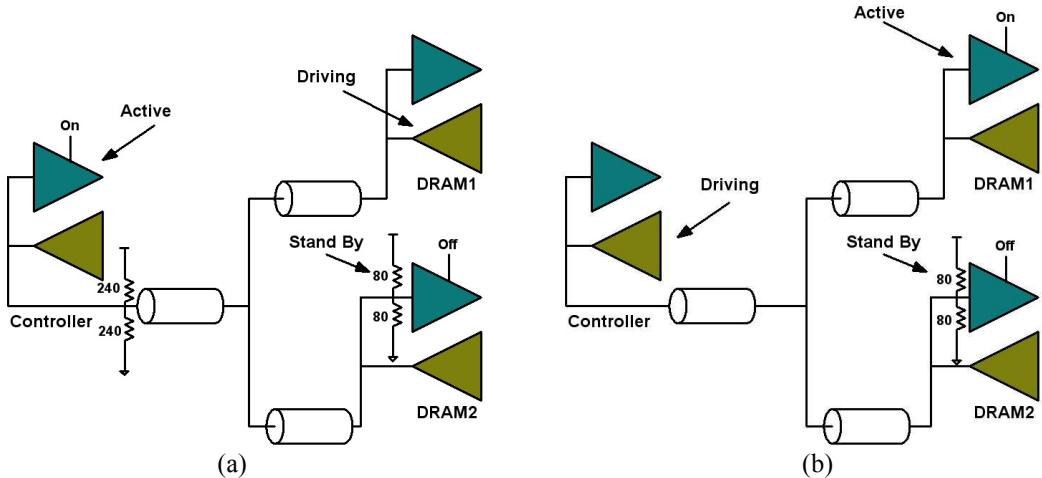

(a) (b)
Figure 5: ODT configuration for (a) memory read, and (b) memory write

A software tool can support this feature by allowing the system designer to specify the IBIS I/O buffer models that are to be used for the data group pins when they are driving, receiving, or in stand-by mode. An IBIS model contains slow/typical/fast VI and TV curves that capture the I/O buffer behavior and account for PVT variations. It is worth mentioning that, in this case, SPICE-like macro models, which are more accurate but slower to simulate than IBIS models, can be considered as another alternative to model I/O buffers.

|  | Driver | Receiver | Standby Receivers |
|---|---|---|---|
| **Controller** | DRVR | RCVR_240 | RCVR_240 |
| **DRAM1** | DRVR | RCVR | ODT_80 |
| **DRAM2** | DRVR | RCVR | ODT_80 |

Table 1: An example showing different I/O and ODT assignments for the two-module system of Figure 5

In an easy-to-use flow, the I/O model assignment process should be performed at the component rather than pin level. Thus, for a given component (e.g., DRAM1 of Figure 5) and for each of its model selectors (specified in its IBIS device model) a driver, receiver, and a stand-by (ODT) model can be selected from the list of models of the corresponding model selector. Hence, all the component pins that share the same model selector use the same I/O and ODT models. Let us assume that the model assignment for the 2-DIMM system of Figure 5 is as shown in Table 1. Based on this assumption, during a write cycle to DRAM1, the driver at the controller side, the active DRAM1 receiver, and the stand-by DRAM2 receiver use models "DRVR", "RCVR", and "ODT_80", respectively. While in a read cycle from DRAM2, the DRAM2 driver, the controller receiver, and the stand-by DRAM1 receiver use models "DRVR", "RCVR_240", and "ODT_80", respectively.

Based on the interface and ODT setup, as well as the signal associations, a software tool can automatically simulate all possible read-write cycle combinations with different ODT and PVT variations. As a result, the design cycle time can be shortened significantly.

## Solution Space Exploration

As we mentioned earlier, unlike the T-branch topology (see Figure 6(a)) that is supported in a DDR2 interface, a DDR3 interface uses a fly-by topology (see Figure 6(b)) to route the address/command/control signals on the various DIMMs. The main design challenge in both cases is to guarantee that the point-to-point strobe/clock signals are in sync with their corresponding address signals at the receiving end. The skew can be minimized in a T-branch topology by balancing the lengths of the stubs. However, in a fly-by topology, the address signals reach the DRAMs at different times (i.e., different delays), so the new read/write-leveling feature must compensate for these delays.

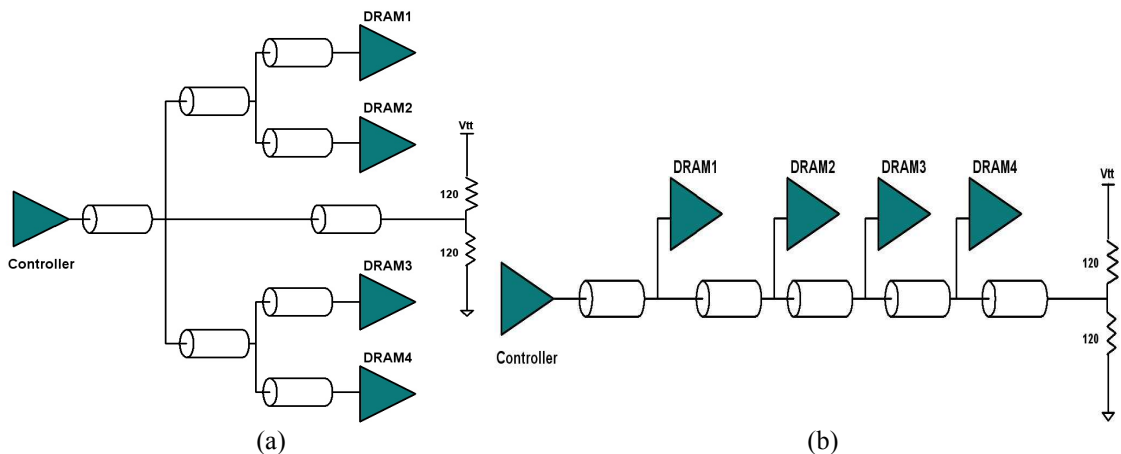

Figure 6: (a) DDR2 T-branch topology, (b) DDR3 fly-by topology

Supporting various types of topologies requires the availability of a robust pre-route SI analysis tool that enables the designer to find optimal system parameters (e.g., stub length, stub impedance, line spacing, and flight time) through solution space exploration. A constraint-driven router can use the developed electrical and physical parameters to try to layout the various signals while meeting the desired timing and noise margins. Moreover, the estimated timing parameters can be used to shift the clock/strobe signals accordingly to compensate for the fly-by topology delays. Hence, a post-layout verification tool should allow assigning Pseudo Random Bit Patterns (PRBP) with different delays to the corresponding signal nets in order to model read/write leveling.

## Timing Measurements and Slew Rate Derating

The setup and hold time margins of a DRAM are based on its response to a 1V/ns input slew rate. Hence, if the measured slew rate of an input signal is not equal 1V/ns, the timing requirements will be different from what the memory vendor is advertising. To address this limitation, slew rate derating tables (see Figure 7) are defined in the DDR2 and DDR3 specifications. The values in these tables depend on the slew rates of the input signals and their corresponding clock/strobe signals. Based on these signals, a setup/hold derating value is retrieved from the corresponding table and then added to the measured setup/hold time in order to verify if the timing constraints are met. The equations to calculate the slew rate of a differential clock/strobe and a data signal are shown in Figures 8 and 9, respectively.

| ΔtDS, ΔDH derating in [ps] AC/DC based [1] | | | | | | | | | | | | | | | | |
|---|---|---|---|---|---|---|---|---|---|---|---|---|---|---|---|---|
| | | DQS, DQS# Differential Slew Rate | | | | | | | | | | | | | | |
| | | 4.0 V/ns | | 3.0 V/ns | | 2.0 V/ns | | 1.8 V/ns | | 1.6 V/ns | | 1.4 V/ns | | 1.2 V/ns | | 1.0 V/ns | |
| | | ΔtDS | ΔtDH | ΔtDS | ΔtDH | ΔtDS | ΔtDH | ΔtDS | ΔtDH | ΔtDS | ΔtDH | ΔtDS | ΔtDH | ΔtDS | ΔtDH | ΔtDS | ΔtDH |
| DQ Slew rate V/ns | 2.0 | 88 | 50 | 88 | 50 | 88 | 50 | - | - | - | - | - | - | - | - | - | - |
| | 1.5 | 59 | 34 | 59 | 34 | 59 | 34 | 67 | 42 | - | - | - | - | - | - | - | - |
| | 1.0 | 0 | 0 | 0 | 0 | 0 | 0 | 8 | 8 | 16 | 16 | - | - | - | - | - | - |
| | 0.9 | - | - | -2 | -4 | -2 | -4 | 6 | 4 | 14 | 12 | 22 | 20 | - | - | - | - |
| | 0.8 | - | - | - | - | -6 | -10 | 2 | -2 | 10 | 6 | 18 | 14 | 26 | 24 | - | - |
| | 0.7 | - | - | - | - | - | - | -3 | -8 | 5 | 0 | 13 | 8 | 21 | 18 | 29 | 34 |
| | 0.6 | - | - | - | - | - | - | - | - | -1 | -10 | 7 | -2 | 15 | 8 | 23 | 24 |
| | 0.5 | - | - | - | - | - | - | - | - | - | - | -11 | -16 | -2 | -6 | 5 | 10 |
| | 0.4 | - | - | - | - | - | - | - | - | - | - | - | - | -30 | -26 | -22 | -10 |

NOTE:
1. Cell contents shaded in red are defined as 'not supported'.

Figure 7: A data derating table example
(This table is taken from JEDEC specification No. 79-3A, Table 73, page 167.)

The derating table of an address signal is not the same as that of a data signal. Moreover, a data/address signal may have two derating tables based on whether its strobe/clock signal is single ended or differential. That contributes to the complexity of the slew rate derating process and necessitates its mitigation to an EDA tool. The tool uses the simulator results to calculate the input setup and hold times as well as the data/strobe nominal- or tangent-line slew rates (see Figure 9). In addition, it allows associating a derating table with a bus. Moreover, it loads the derating table and looks up a setup or hold derating value based on the measured slew rates of the input and strobe signals. Furthermore, it adds the setup and hold derating values to the setup and hold times, respectively.

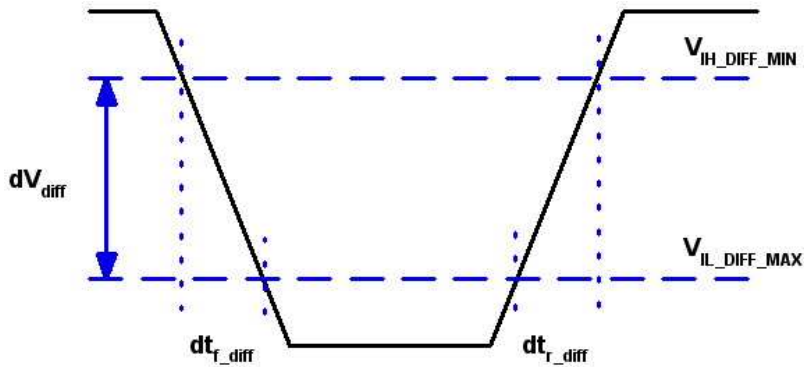

Figure 8: Slew rate for differential input voltage (i.e., strobe and strobe# as well as clock and clock#): (a) for falling edge = $dV_{diff}/dt_{f\_diff}$, and (b) for rising edge = $dV_{diff}/dt_{r\_diff}$

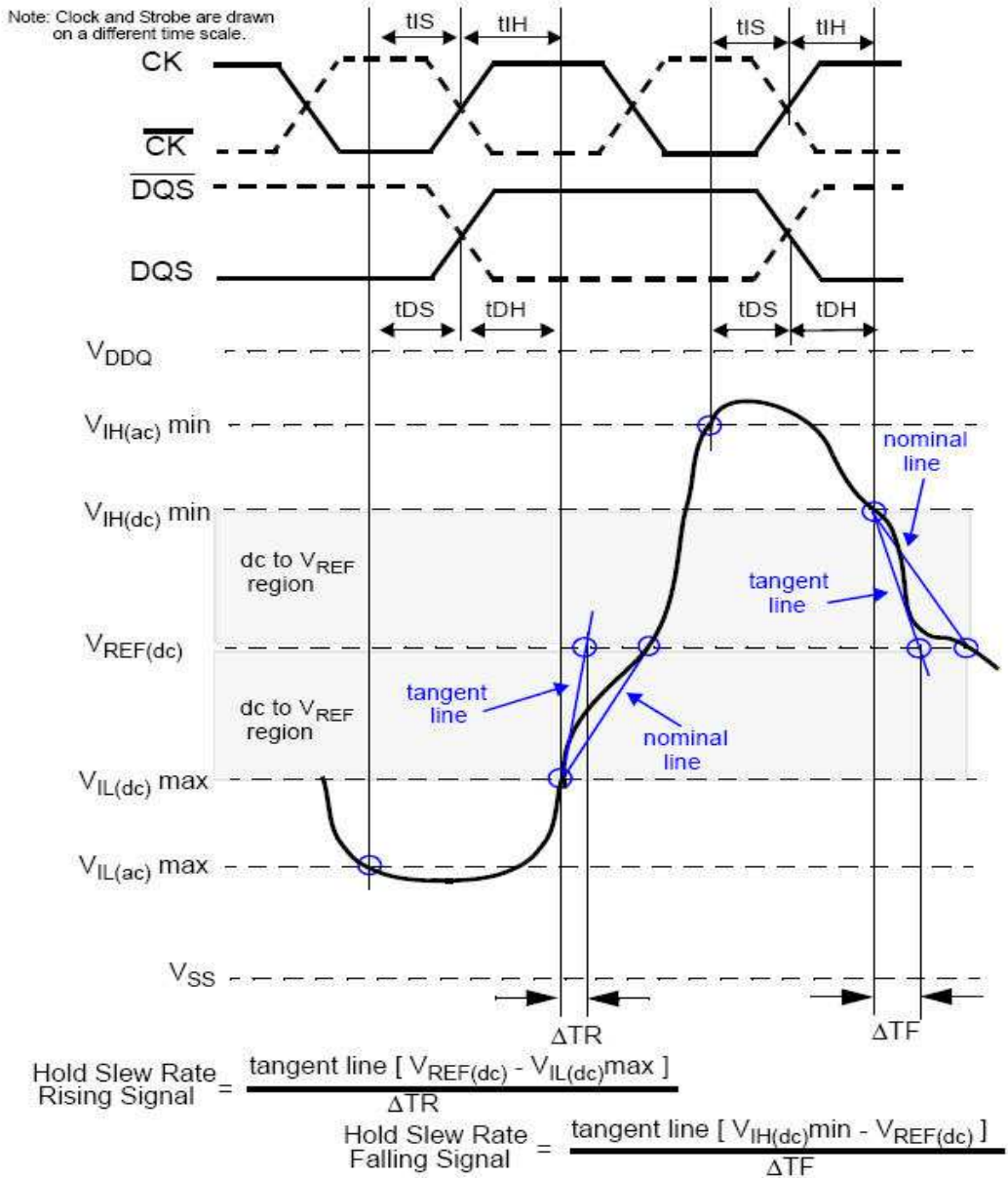

Figure 9: Setup and hold slew rate calculations for a data signal
(This figure is taken from JEDEC specification No. 79-3A, Figure 108, page 171.)

## SI Analysis and Measurements

After setting up and laying out a DDR3 interface, it should be validated to see if it meets the desired timing and noise margins. Since verifying a DDR3 design is a challenging and complex process, the tool used to accomplish this phase of the design must have the following features: (1) support all DDR3 features, (2) produce accurate results relative to measurements, (3) easy to use, and (4) automate the process as much as possible. Hence, the tool must simulate automatically all possible driver/receiver combinations for all ODT and I/O PVT variations. In addition, it should calculate the timing (e.g., setup and hold times) and noise (e.g., overshoot, noise margin) information based on the AC and DC thresholds (see Figure 10(a)) defined in the SSTL specification. Moreover, it needs to support derating tables and slew rate derating. Also, it should generate manageable reports and display eye diagrams with a data valid window overlaid on top (see Figure 10(a)) in order to identify and then reroute the net(s) that violate any of the design rules.

The tool should also support two types of SI analysis: reflection and comprehensive. The reflection analysis is needed to study the discontinuity effects on the SI. So, in this case, each bit (data, address, clock, etc.) is simulated independently from its neighbors. On the other hand, the discontinuity (reflection), coupling (xtalk), and SSN effects are all taken into consideration in a comprehensive analysis. Hence, in this scenario, each victim net is extracted along with its neighboring net(s) to model xtalk. Moreover, to account for the SSN effects, the power/ground planes should be modeled and connected to the corresponding power rails. In both scenarios, all the passive structures (e.g. wire bonds, solder balls, interconnects, delay lines, vias, power planes, connector, as shown in Figure 10(b)) between the driver(s) and receiver(s) (i.e. from die to die) should be extracted and modeled in order to produce accurate simulation results. Since the data rate of a DDR3 interface can be as much as 1600Mbps, it requires the validity of the passive models within a multi-GHz frequency range. Thus, a 3D field solver, rather than a quasi-static solver, is needed to generate these high bandwidth models.

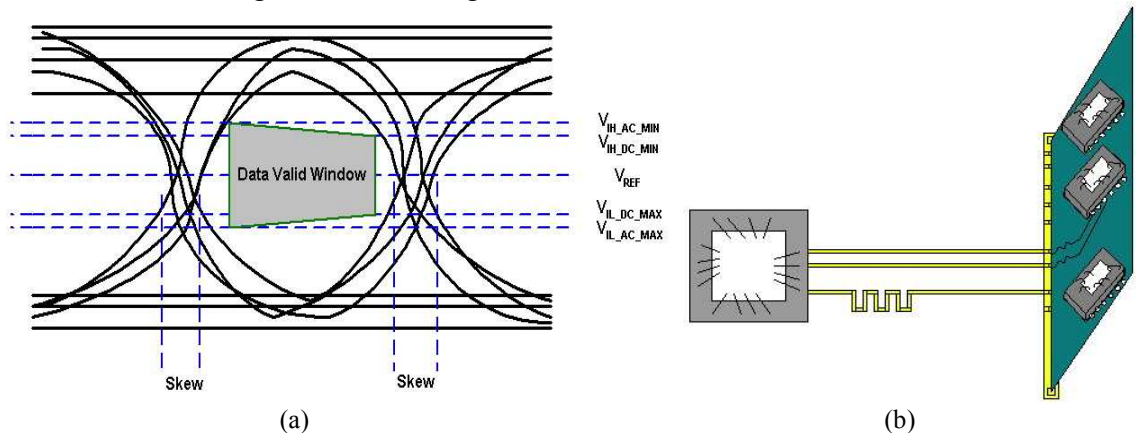

(a) (b)
Figure 10: (a) an eye diagram example, (b) typical passive structures on a DDR3 system

## Summary

In this paper, we presented a methodology for tackling the various DDR3 interface design challenges. Methodology includes pre-layout exploration of various combinations of conditions to develop constraints that drive the PCB layout process. It shows how to automate the various read-write cycle combinations with different ODTs and PVT variations to shorten the time needed for post-layout verification. In addition, it shows how to ensure that the setup and hold timing margins are not eroded by reflection and crosstalk noise in the byte lanes. Moreover, it addresses features such as slew rate derating as well as read and write leveling to compensate for the fly-by topology delays.